\documentclass[twocolumn]{autart}

\usepackage{amsmath,amssymb,amsfonts}
\usepackage{graphicx,color}  
\usepackage{enumitem}
\usepackage[normalem]{ulem}
\usepackage{babel}
\usepackage{todonotes}

\usepackage{tikz}                         
\usepackage{url}
\usepackage{hyperref}

\newcommand{\IN}{\text{in}}
\newcommand{\OUT}{\text{out}}
\newcommand{\LEAK}{\text{leak}}

\begin{document}

\begin{frontmatter}

\title{On well-posedness of the leak localization problem \\in parallel pipe networks\thanksref{footnoteinfo}} 

\thanks[footnoteinfo]{This paper was not presented at any IFAC 
meeting. The work was supported by the Digital Futures project Democritus. Corresponding author Victor Moln\"o.}

\author[KTH,DF]{Victor Moln\"o}\ead{vmolno@kth.se},    
\author[KTH,DF]{Henrik Sandberg}\ead{hsan@kth.se}      

\address[KTH]{Division of Decision and Control Systems, KTH Royal Insitute of Technology, Stockholm, Sweden}
\address[DF]{Digital Futures, KTH Royal Insitute of Technology, Stockholm, Sweden}


\begin{abstract}                          
With the advent of integrated sensor technology (smart flow meters and pressure sensors), various new numerical algorithms for leak localization (a core element of water distribution system operation) have been developed. However, there is a lack of theory regarding the limitations of leak localization. In this work, we contribute to the development of such a theory by introducing an example water network structure with parallel pipes that is tractable for analytical treatment. We define the leak localization problem for this structure and show how many sensors and what conditions are needed for the well-posedness of the problem. We present a formula for the leak position as a function of measurements from these sensors. However, we also highlight the risk of finding false but plausible leak positions in the multiple pipes. We try to answer the questions of how and when the leaking pipe can be isolated. In particular, we show that nonlinearities in the pipes' head loss functions are essential for the well-posedness of the isolation problem. We propose procedures to get around the pitfall of multiple plausible leak positions.
\end{abstract}

\end{frontmatter}

\section{Introduction}
Leakage in water networks is a worldwide major societal concern. According to estimates, about a hundred billion cubic meters leak out annually, accounting for approximately 30\% of the input volume~\cite{leakstatistics}. The lost water is a problem in itself when availability is scarce, but leakage also results in wasted resources for water treatment, pumping, etc. In some cases, like in Cape Town, 2018, the loss is even more noticeable as leakages contribute to residents partially or wholly losing access to drinking water~\cite{capetown}. Furthermore, water escaping from leaking pipes may undermine and damage infrastructure, and the leak hole may provide an access point for pollutants~\cite{boston,pollutant2}.

SCADA systems with integrated sensors are used to monitor water distribution systems, and an important part of the monitoring lies in the detection and localization of leakages. Examples of systems for these tasks are described in~\cite{SysDetLoc,rathore+2022}. However, there is no obvious best way to utilize the sensor measurements. There are rather many alternative algorithms that utilize various sensor data and assumptions. For further examples, see the survey~\cite{Survey} and references therein. Case studies, for instance, in~\cite{SysDetLoc,rathore+2022}, and simulation benchmark performance tests, such as the BattLeDIM competition~\cite{BattLeDIM}, give practical indications of effective solutions. However, much less work is done on the theoretical guarantees and foundations of leak localization. The fundamental question of when a leak is possible to unambiguously localize, in terms of network structure and minimal necessary sensor information, appears to be open. In this paper, we answer this theoretical question in a parallel pipe configuration, which is analytically tractable.

Our analysis relies on traditional, well-studied steady-state water system models. In~1936, the Hardy Cross method was introduced to compute a hydraulic state solution~\cite{HardyCross}. In 1956, Birkhoff and Diaz published results regarding the existence and uniqueness of this hydraulic state solution, given a nonlinear flow network, for certain boundary conditions~\cite{Birkhoff}. Since then, more computationally efficient hydraulic state solution methods have been developed (see \cite{TodiniPilati} for an important example and \cite{WDNhistory} for more historical notes). This development has led to widely used simulation tools such as EPANET~\cite{epanet}. But despite all of these efforts with computational models, there is still a need for more theory regarding the localizability of leaks. In particular, such a theory can help identify the minimum measurement resources required to identify leaks (the observability problem) and hydraulic states where leaks are possible or easier to isolate (leaking pipe isolation and active fault detection problems). Such theory can guide us in designing especially challenging leak localization problems, cf.~BattleDIM~\cite{BattLeDIM}, and could be used to help design future networks that are easier to maintain. The leak localization theory we have in mind is different from existing observability analysis of water distribution systems such as~\cite{saraidiaz}, which deals with hydraulic state estimation. At this point, it is also interesting to make a connection to recent developments in electric power systems and secure control systems. Energy Management Systems and state estimators are routinely used to operate power systems to optimally use infrastructure resources and increase fault resilience. Observability analysis, similar to analysis developed here for water systems, has been used to identify security flaws and weak spots in state estimators, see~\cite{liu+11,teixeira+15}. As more online monitoring and control applications are introduced in water systems, similar security problems should be anticipated in the water domain.

It should be noted that there exists theoretical work with guarantees for leak localization, for example,~\cite{norge1,norge2,norge3}, based on dynamical PDE models. The first two of these, \cite{norge1} and \cite{norge2}, analyze single pipes and branched networks. The third paper,~\cite{norge3}, considers leak diagnosis in a ring-shaped network structure (a mesh or a loop), similar to our model. However,~\cite{norge3} assumes an auxiliary flow sensor in an individual pipe. The main difference between our paper and \cite{norge1,norge2,norge3} is that we do not assume the high-frequency sampling required to analyze pressure and flow transients, and, therefore, we work with the previously mentioned steady-state models.

In this paper, we generalize the single pipe leak localization problem of~\cite{lindstrom2022leakage} to parallel pipe networks (Problem~\ref{problem}). 
This network structure is simple enough to allow for analytical treatment, yet we identify several conditions under which \emph{the leak localization problem has no unique solution and is ill-posed}. Unlike most other works, we do not restrict the possible leak position to a pre-determined, finite set of junctions or consumer locations. Rather, we consider the possibility of a leak anywhere along any of the pipes and single out possible locations consistent with the available sensor data. We make the following specific contributions:

\begin{enumerate}[label = \arabic*)]
    \item We prove that a particular set of sensors is needed to solve Problem~\ref{problem} (Theorem~\ref{thm:necessearydata}). The authors of~\cite{steliosmodelinvalidation} recognize that the leak localization problem in larger networks is almost always under-determined. The paper~\cite{optimalplacement} presents a heuristic approach to optimal sensor placement for leak localization. However, these works do not delve into the theoretical lower bound on the required number of sensors, which we do.
    \item We show that for Problem~\ref{problem} to be well-posed, measurements in a single hydraulic state is not enough (Proposition~\ref{prop:newprop}), in contrast to the single pipe problem in~\cite{lindstrom2022leakage}. 
    \item We prove that two different hydraulic states, satisfying certain conditions (Theorem~\ref{thm:dh0cond}), are sufficient to solve Problem~\ref{problem}. Conversely, we prove the existence of so-called \emph{confusion flows} where it is \emph{impossible to decide which pipe is leaking} (Proposition~\ref{prop:confusion_flow}), and the problem is then ill-posed.
    \item \emph{We present two scenarios where the leaking pipe cannot be uniquely determined, despite measurements in any number of states (Theorem~\ref{thm:identical} and Theorem~\ref{thm:impossiblelinear}), and Problem~\ref{problem} is then inherently ill-posed.}
   However, by introducing side information about the leak model, the impossibility result of Theorem~\ref{thm:impossiblelinear} can be circumvented (Theorem \ref{thm:leakfnccontradictionlinear}).
\end{enumerate}

In Section~\ref{sec:model}, we introduce the model of our pipe structure and present a residual function equivalent to its unique solvability through measurements. We also define our leak localization problem. In~Section~\ref{sec:localization}, we show that sensor pressure and flow measurements in the junctions are necessary and sufficient to calculate a leakage position. Here we see also, however, that we can calculate one plausible leakage position per pipe. In Section~\ref{sec:experiment}, we show how to manipulate the system to isolate the leaking pipe using multiple measurements. In doing so, we identify leak cases that, under some conditions, are indistinguishable. Finally, in Section~\ref{sec:impossible}, we show cases where the leak localization is inherently impossible using measurements and infrastructure models alone. One of the cases is solvable by introducing auxiliary leak model characteristics.

\section{Parallel pipes model and leak localization problem}
\label{sec:model}

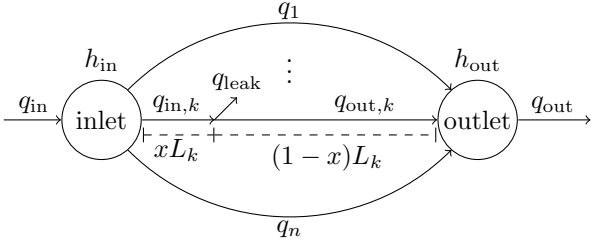
\begin{figure}[tb]
    \centering
    \begin{tikzpicture}
    \node at (0,0) {inlet};
    \node at (5,0) {outlet};
    \node at (-0.9,0.2) {$q_{\IN}$};
    \node at (6,0.2) {$q_{\OUT}$};
    \node at (0,0.8) {$h_{\IN}$};
    \node at (5,0.8) {$h_{\OUT}$};
    \draw[black] (0,0) circle (15pt);
    \draw[black] (5,0) circle (15pt);
    \draw[->] (-1.3,0) -- (-0.54,0);
    \draw[->] (5.54,0) -- (6.5,0);
    \path[->,out = 45, in = 135] (0.35,0.4) edge (4.65,0.4);
    \path[->,out = -45, in = -135] (0.35,-0.4) edge (4.65,-0.4);
    \draw[->] (0.54,0) -- (1.5,0);
    \draw[->] (1.5,0) -- (4.46,0);
    \draw[->] (1.5,0) -- (1.8,0.3);
    \node at (2.5, 1.45) {$q_{1}$};
    \node at (2.5, -1.45) {$q_{n}$};
    \node at (1, 0.2) {$q_{\IN,k}$};
    \node at (3.5, 0.2) {$q_{\OUT,k}$};
    \draw[|-|,dashed] (0.55 , -0.2) -- (1.5,-0.2);
    \draw[-|,dashed] (1.5 , -0.2) -- (4.45,-0.2);
    \node at (1,-0.4) {$xL_k$};
    \node at (3,-0.5) {$(1-x)L_k$};
    \node at (1.8,0.5) {$q_{\LEAK}$};
    \node at (2.5,0.75) {$\vdots$};
    \end{tikzpicture}
    \caption{Schematic view of a network of $n$ parallel pipes, where pipe $k$ is leaking.}
    \label{fig:model}
\end{figure}

\begin{figure}[tb]
    \centering \includegraphics[width=0.5\hsize]{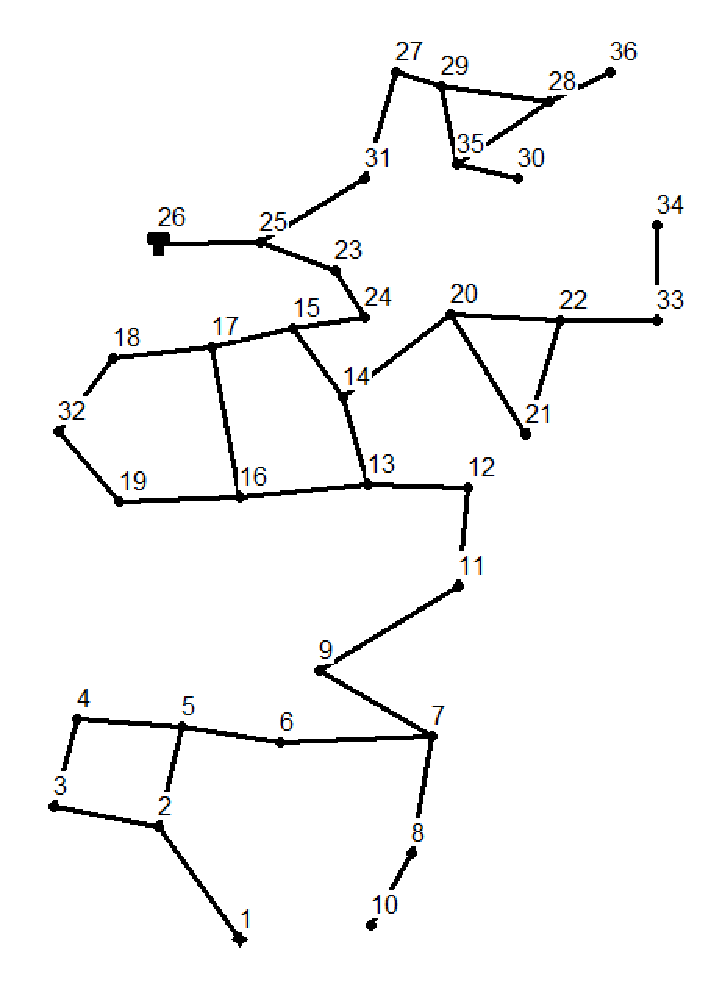}
    \caption{A water network with some parallel pipe structures. (EPANET~\cite{epanet} Example Network~2, available at \href{https://github.com/OpenWaterAnalytics/epanet-example-networks}{github.com/OpenWaterAnalytics/epanet-example-networks}.)
    }
    \label{fig:EPANET-example}
\end{figure}

In this work, we consider a subnetwork in a (possibly much larger) water network, with two junctions connected by $n$ parallel pipes, as seen in Fig.~\ref{fig:model}. Parallel pipes between a pair of junctions introduce redundancy and allow for alternative flow paths in case of failures, e.g., due to leaks. Parallel pipes can also help balance pressures, reducing the risk of pollution due to stagnation. Regardless of the reason for the parallel pipes, we describe a common structure in water networks. The network in Fig.~\ref{fig:EPANET-example} from EPANET~\cite{epanet} has several instances of two parallel pipes ($n=2$), between junctions 2 and 5, 16 and 17; and 20 and 22, for example. 

At one junction of our mesh, the inlet, there is an inflow of water $q_{\IN}$ [volume units per time unit]. At the other junction, the outlet, there is an outflow $q_{\OUT}$. There are \emph{hydraulic heads} of $h_{\IN}$ and $h_{\OUT}$ [length units] at the inlet and outlet, respectively. Hydraulic head is the sum of the \emph{pressure head}, which is the height of a water column exerting the pressure of the water, and the \emph{elevation head}, which is the elevation of the pressure measurement point with respect to a system-wide reference point. In practice, pressure sensors will give readings of pressure head, but this is easy to translate to hydraulic head \cite{encyclopedia}. We will refer to hydraulic head simply as \emph{head}.

We assume that $h_{\IN}, h_{\OUT}, q_{\IN}$, and $q_{\OUT}$ are measured by sensors installed at the inlet and outlet. This is the maximal of all point-wise pressure and flow sensor configurations, restricted to installation at the junctions. We will see in Theorem~\ref{thm:necessearydata}, Section~\ref{sec:localization} that for leak localization, this maximal sensor configuration is, in fact, necessary. Furthermore, due to a corresponding sufficiency result, Proposition~\ref{prop:newprop}, we can solve for possible leak locations analytically for the model in Fig.~\ref{fig:model}. 

\begin{rem}
In Fig.~\ref{fig:EPANET-example}, we see nodes inside some of the parallel pipe interconnections; nodes~3 and 4, for example, where water extraction may occur. In the following, we assume there is no such extraction during fault localization, for simplicity. Note, however, that our results can be generalized to non-zero extraction at such nodes through the use of consumption meters.
\end{rem}

Our scenario involves one leak, $q_{\LEAK}$, located $xL_k$ length units along pipe $k$ downstream from the inlet. Here $L_k$ is the full length of pipe $k$. We call $x\in(0,1)$ the \emph{relative} leak position.

We assume that the head and the flow in the pipes are always in a  steady (hydraulic) state, and related through~$-\frac{dh_i(z_i)}{dz_i} = U_i(q_i)$, where $z_i$ is the relative position in pipe $i$, $h_i(z_i)$ is the head at said relative position, $q_i$ is the flow through pipe $i$, and~$U_i$ is the \emph{head loss function} of pipe $i$. We assume all head loss functions $U_i$ are strictly increasing in $q_i$ and therefore \emph{invertible}.

We assume that the water network operator has access to accurate models of all $U_i$.
\begin{rem}
    Complete knowledge of all $U_i$ is a strong assumption. The same holds for noiseless measurements, which we also assume. In the literature, usually, both of these assumptions are relaxed somehow. For example, in~\cite{steliosintervaldetection}, head loss function parameters and sensor errors are assumed to fall within specified uncertainty intervals. The uncertainty then carries over to the leak localization. In an ongoing work of ours, we focus on simultaneously estimating the a~priori unknown head loss function parameters and localizing the leak using noisy sensor data. However, the analysis in the current study still carries value since if the problem is shown to be hard under strong assumptions, the problem remains hard also under weaker assumptions. 
\end{rem}

We assume uniform flow along the length of each pipe, so the total head loss between the inlet and the outlet is $\Delta h := h_{\IN}-h_{\OUT} = U_i(q_i)$. The leaking pipe $k$ is an exception, where the flow  $q_{\IN,k}$ upstream of the leak is not the same as the flow $q_{\OUT,k}$ downstream. Writing down the total head loss $\Delta h$ for each pipe, as well as the total flow through the $n$-pipe network, we obtain the following network model:
\begin{align}
    \Delta h &= U_i (q_i), \quad i \neq k, \label{eq:dHnonleak}\\
    \Delta h& = xU_k(q_{\IN,k}) +  (1-x)U_k(q_{\OUT,k}), \label{eq:dHleak}\\
    \label{eq:qin}
    q_{\IN} &= q_{\IN,k} + \sum_{i\neq k} q_{i},\\
    q_{\OUT} & = q_{\OUT,k} + \sum_{i \neq k} q_{i}. \label{eq:qout}
\end{align}
The equations (\ref{eq:dHnonleak})--(\ref{eq:qout}) constitute all physical relations we will use for leak localization up until Section~\ref{sec:leakfunction}, where we assume an auxiliary leak function, modeling the characteristics of the leak itself. Leak localization for the described model means to solve Problem~\ref{problem}.
\begin{prob}
\label{problem}
Given measurements of $h_{\IN}, h_{\OUT}, q_{\IN}$, and $q_{\OUT}$, isolate the leaking pipe~$k$ and find the leak location~$x$.
\end{prob}

\section{Preliminary results on leak candidates}
\label{sec:localization}

In this section, we present some preliminary results, showing that a \emph{data point}~$(q_{\IN},q_{\OUT},h_{\IN},h_{\OUT})$ consisting of simultaneous measurements (one per variable), corresponds to exactly one leak position per pipe. We call these $n$ leak positions \emph{leak candidates}.

We let the flow admittance function $G_{-j}(\Delta h) := \sum_{i \neq j} U_i^{-1}(\Delta h)$,
denote the total flow through all pipes except pipe $j$ when these are non-leaking and the total head loss is $\Delta h$. We take the residual function
\begin{multline*}
    r_j(x_j,\Delta h, q_{\IN},q_{\OUT}) 
 :=\Delta h -x_jU_j(q_{\IN}-G_{-j}(\Delta h))  
    \\-(1-x_j)U_j(q_{\OUT}-G_{-j}(\Delta h)), \label{eq:r}
\end{multline*}
as a measure of the discrepancy between the true solution to (\ref{eq:dHnonleak})--(\ref{eq:qout}) and the solution if the leak was in the pipe $j$ in the relative position $x_j$. According to (\ref{eq:dHleak}), $x$ in pipe $k$ is the true leak location, so $r_k(x,\Delta h, q_{\IN}, q_{\OUT}) \equiv 0$. Lemma~\ref{lem:r} relates the residual function to leak position evaluation.

\begin{lem}    
    \label{lem:r}
    \begin{enumerate}[label = \arabic*)]
        \item ($\Delta h, q_{\IN}, q_{\OUT}, q_{\IN,k}, q_{\OUT,k}, \{q_i\}_{i \neq k})$ solve the network model (\ref{eq:dHnonleak})--(\ref{eq:qout}) only if 
        \begin{equation}
            r_k(x,\Delta h,q_{\IN},q_{\OUT}) = 0.
        \label{eq:resid_test}
        \end{equation}
        \item If $r_k(x,\Delta h,q_{\IN},q_{\OUT}) = 0$, then ($\Delta h, q_{\IN}, q_{\OUT}, q_{\IN,k},$ $q_{\OUT,k}, \{q_i\}_{i \neq k}$) where $q_i = U^{-1}_i(\Delta h)$, $i \neq k$, $q_{\IN,k} = q_{\IN}-G_{-k}(\Delta h)$ and $q_{\OUT, k} = q_{\OUT}- G_{-k}(\Delta h)$ solve (\ref{eq:dHnonleak})--(\ref{eq:qout}).
    \end{enumerate}
\end{lem}

\begin{pf}
\begin{enumerate}[label = \arabic*)]
    \item If $\Delta h, q_{\IN}, q_{\OUT}, q_{\IN,k}, q_{\OUT,k}, \{q_i\}_{i \neq k} $ solve the model equations (\ref{eq:dHnonleak})--(\ref{eq:qout}), then by (\ref{eq:dHnonleak}), $q_i = U_i^{-1}(\Delta h)$, and so by (\ref{eq:qin}) and (\ref{eq:qout}), $q_{\IN,k} = q_{\IN}-G_{-k}(\Delta h)$ and $q_{\OUT,k} = q_{\OUT}-G_{-k}(\Delta h)$. By (\ref{eq:dHleak}), $r_k(x, \Delta h, q_{\IN}, q_{\OUT}) = 0$.
    \item By construction, $\{q_i\}_{i \neq k}$ solve (\ref{eq:dHnonleak}). Similarly, $q_{\IN,k}$ and $q_{\OUT,k}$ solve (\ref{eq:qin}) and (\ref{eq:qout}). Plugging in $q_{\IN,k}$ and $q_{\OUT,k}$ for $q_{\IN}- G_{-k}(\Delta h)$ and $q_{\OUT}-G_{-k}(\Delta h)$ in $r_k(x, \Delta h, q_{\IN}, q_{\OUT}) = 0$, we solve (\ref{eq:dHleak}). \qed
\end{enumerate}
\end{pf}

\begin{rem}
    Lemma \ref{lem:r} hints that checking estimates of $x$ against the model (\ref{eq:dHnonleak})--(\ref{eq:qout}) can be reduced to a residual test~(\ref{eq:resid_test}). Many of the new integrated sensor-based leak localization algorithms follow a similar procedure. First, a leak position is assumed. Then the system of governing physical equations is solved, based on a subset of the sensor measurements. The remaining sensor measurements are used to evaluate the assumed leak position.
\end{rem}

\begin{prop}
\label{prop:newprop}
For every data point $(h_{\IN}, h_{\OUT}, q_{\IN}, q_{\OUT})$ such that $q_{\IN} \neq q_{\OUT}$, there is exactly one $x_j \in (0,1)$ for each pipe $j= 1, \dots, n$ for which $r_j(x_j, \Delta h, q_{\IN}, q_{\OUT}) = 0$ given by
\begin{align}
    \label{eq:ludvig}
    x_j = \dfrac{\Delta h-U_j(q_{\OUT}- G_{-j}(\Delta h))}{U_j(q_{\IN}- G_{-j}(\Delta h))-U_j(q_{\OUT}- G_{-j}(\Delta h))}.
\end{align}
For the truly leaking pipe $k$, we have $x_k = x$.
\end{prop}
\begin{pf}
When $q_{\IN} \neq q_{\OUT}$, the residual~$r_j(x_j, \Delta h, q_{\IN},q_{\OUT})$ is linear in $x_j$ (with non-zero slope). We show that $r_j(0, \Delta h, q_{\IN},q_{\OUT})>0>r_j(1, \Delta h, q_{\IN},q_{\OUT})$. Thus there is a unique $x_j \in (0,1)$ (given by (\ref{eq:ludvig})) that solves $r_j(x_j, \Delta_h , q_{\IN},q_{\OUT}) = 0$. Showing the inequalities: 
\begin{align*}
   r_j(0,& \Delta h, q_{\IN},q_{\OUT})
        \\& = \Delta h - U_j(q_{\OUT}-G_{-j}(\Delta h)) 
        \\
          &= \Delta h-U_j(q_{\OUT,k} + G_{-k}(\Delta h)- G_{-j}(\Delta h))
          \\
           & = \Delta h-U_j(q_{\OUT, k}-q_{k} + U_j^{-1}(\Delta h))
           \\
           & > \Delta h -U_j(U_j^{-1}(\Delta h)) = \Delta h - \Delta h = 0.
           \end{align*}
           Similarly, 
           \begin{align*}
           r_j(1, &\Delta h, q_{\IN},q_{\OUT}) 
           \\& = \Delta h-U_j(q_{\IN, k}-q_{k} + U_j^{-1}(\Delta h))
           \\
           & < \Delta h-U_j(U_j^{-1}(\Delta h)) = 0. 
\end{align*}
Here, $q_k = U_k^{-1}(\Delta h)$ is the flow that would pass through pipe $k$ under a  total head loss of $\Delta h$, assuming pipe $k$ was not broken. We have used that $q_{\IN,k} > q_k > q_{\OUT,k}$, however the same result follows from $q_{\IN,k} < q_k < q_{\OUT,k}$. To show that we have either of these cases, we know that~$xU_k(q_{\IN,k}) + (1-x)U_k(q_{\OUT,k}) = \Delta h = U_k(q_{k})$. That is, either~$U(q_{\IN,k}) > U(q_k) >U(q_{\OUT,k})$ or~$U(q_{\IN,k}) < U(q_k) <U(q_{\OUT,k})$. Because~$U_k$ is strictly increasing, either~$q_{\IN,k} > q_{k} > q_{\OUT,k}$, or~$q_{\OUT,k} > q_{k} > q_{\IN,k}$. \qed
\end{pf}

\begin{rem}
Proposition \ref{prop:newprop} is an adaptation of Theorem~1 from \cite{lindstrom2022leakage}. The work~\cite{lindstrom2022leakage} deals with one pipe, i.e., the special case $n = 1$. The formula (\ref{eq:ludvig}) is also similar to the calculations in~\cite{sameformula}.
\end{rem}

According to Proposition~\ref{prop:newprop}, the maximal measurement selection ($h_{\IN}, h_{\OUT}, q_{\IN},q_{\OUT}$) is sufficient to determine $x$, assuming we know the leaking pipe $k$. It turns out that the selection is also necessary.
\begin{thm}
\label{thm:necessearydata}
   For any $x_j \in (0,1)$ and any values of three elements in the data point $(h_{\IN},h_{\OUT},q_{\IN},q_{\OUT})$, there is a unique value of the fourth element which combined solve $r_j(x_j, \Delta h, q_{\IN},q_{\OUT}) = 0$.
\end{thm}
\begin{pf}
    We prove for the different selections of elements, that we can find a unique value for the fourth element to solve $r_j(x_j, \Delta h, q_{\IN}, q_{\OUT}) = 0$, for any $j$.
    \begin{itemize}
        \item Missing $q_{\IN}$ (missing $q_{\OUT}$ follows analogously):
            As $x_j\in (0,1)$, the residual $r_j(x_j, \Delta h, q_{\IN}, q_{\OUT})$ is continuous, strictly decreasing and unbounded in $q_{\IN}$. Thus there will be a unique value $q_{\IN}$ such that $r_j(x_j,\Delta h, q_{\IN}, q_{\OUT})=0$. 
        \item Missing $h_{\IN}$ (missing $h_{\OUT}$ follows analogously):
            $\Delta h$ is linear in $h_{\IN}$. Furthermore $U_i^{-1}(\Delta h)$ is continuous, increasing and unbounded in $\Delta h$. Therefore so is $G_{-j}(\Delta h)$. Since $x_j, 1-x_j \in (0,1)$, and $U_j$ is continuous, strictly increasing, and unbounded, we conclude that $r_j(x_j, \Delta h , q_{\IN}, q_{\OUT})$ is continuous, increasing and unbounded in $\Delta h$. Thus there is a $h_{\IN}$ such that $r_j(x_j, h_{\IN}-h_{\OUT}, q_{\IN}, q_{\OUT}) = 0$. \qed
    \end{itemize}
\end{pf}
According to Theorem~\ref{thm:necessearydata}, we can not uniquely solve for~$x$ given only three measurements. Naturally not given only one or two measurements, either. The assumed sensor configuration is indeed necessary for the well-posedness of the leak localization Problem~\ref{problem}. If any of these sensors are missing, additional assumptions have to be made, for instance of consumption models or pseudo-measurements.

With Proposition~\ref{prop:newprop} and Theorem~\ref{thm:necessearydata} we have seen that a data point $(h_{\IN}, h_{\OUT}, q_{\IN}, q_{\OUT})$ gives us one unique leak position per pipe. The rest of the paper deals with the process of eliminating the leak positions in pipes~$j\neq~k$, using more data points. A first attempt is given in Example~\ref{exmp:differentx}.

\begin{exmp}
\label{exmp:differentx}
   Fig.~\ref{fig:differentx}. shows $x_j$ estimations for a network with $n = 3$ parallel pipes. Here, we have calculated the relative leak positions $x_j$ for $j = 1, 2,3$, and $N = 100$ data points in different hydraulic states. The pipe is simulated with a pressure-dependent leakage. However, we use only~(\ref{eq:ludvig}) to derive the estimations, i.e., we do not rely on knowledge of the leak model. As we see, the estimations of $x_1$ and $x_3$ differ for different data points. Only $x_2$ is constant. Thus, we conclude that it must be pipe $k = 2$ that is leaking, in relative position $x = 0.3$. Here we have used $U_i(q) = c_i|q|q$, with $c_i \in \{0.05, 0.1,0.2\}$.
\begin{figure}
    \centering
    \includegraphics[scale = 0.5]{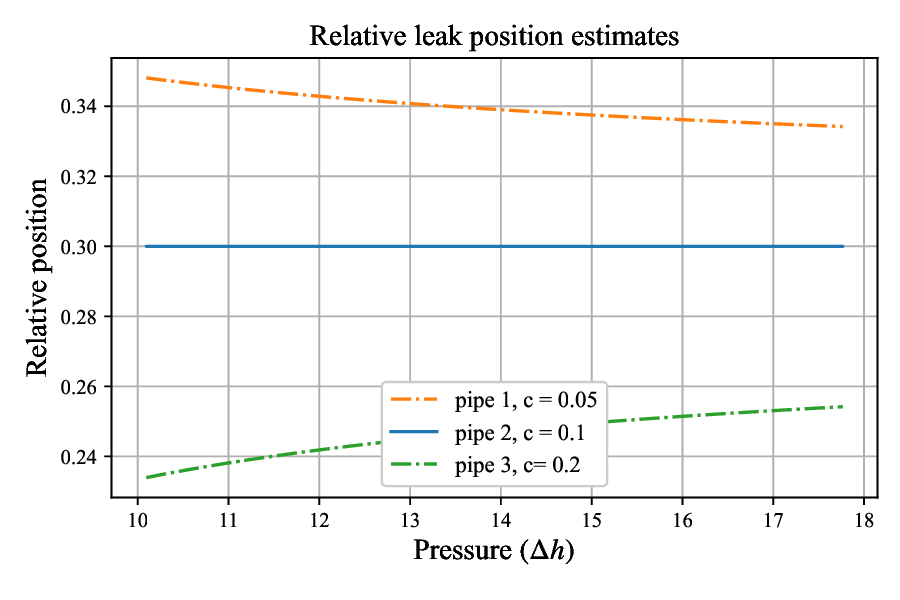}
    \caption{Estimates of leak position $x$  for a network with three parallel pipes. The leak should not move as we vary the pressure, and we conclude that pipe~2 is the leaking pipe.}
    \label{fig:differentx}
\end{figure} 
\end{exmp}

\section{Sufficient conditions for leak isolation}
\label{sec:experiment}

From Proposition~\ref{prop:newprop}, we know that from a single data point $(h_{\IN}, h_{\OUT}, q_\IN,q_\OUT)$ (the nominal state) we can determine $n$ candidate leak positions $x_j$, such that $r_j(x_j,\Delta h, q_\IN,q_\OUT)=0$. In this section, we will provide conditions under which a small perturbation $(dh_{\IN}, dh_{\OUT},dq_\IN,dq_\OUT)$ to the data point is sufficient to isolate the leaking pipe~$k$, and refute the other candidate locations~$x_i$. That is, $r_k(x,\Delta h+dh, q_\IN+dq_\IN,q_\OUT+dq_\OUT)=0$, but $r_i(x_i,\Delta h+dh, q_\IN+dq_\IN,q_\OUT+dq_\OUT)\neq 0$, $i\neq k$, where $dh = dh_{\IN}-dh_{\OUT}$.
We saw in Example~\ref{exmp:differentx} that this is possible, but here we provide an analysis of when, and when not, such perturbations exist.

It turns out that the analysis is easier in a transformed, but to $r_j$ equivalent, residual function, which we call $\bar r_j$ (see Remark~\ref{rem:barrj}). It is defined as,
\begin{equation}
    \bar r_j(x_j,\Delta h, q_\IN,q_\OUT) := q_\OUT - \hat q_{\OUT}(j,x_j,\Delta h, q_\IN) 
\end{equation}
where
\begin{multline*}
     \hat q_{\OUT}(j,x_j,\Delta h, q_\IN)  \\ := U_j^{-1}\left[\frac{\Delta h}{1-x_j} -\frac{x_j}{1-x_j}U_j[q_\IN -G_{-j}(\Delta h)] \right] \\+ G_{-j}(\Delta h).
\end{multline*}
An interpretation of $\bar r_j$ is that the actual outflow $q_\OUT$ is compared to the estimated outflow $\hat q_\OUT$ using the other data $(h_{\IN}, h_{\OUT},q_\IN)$ and \emph{under the assumption that the leak is in pipe~$j$, at position $x_j$.} This interpretation follows from simple manipulation of~\eqref{eq:dHnonleak}--\eqref{eq:qout}, and in particular $q_\OUT\equiv \hat q_\OUT(k,x,\Delta h,q_\OUT)$. The equivalence between $r_j$ and $\bar r_j$ can be stated as follows.
\begin{lem}\label{lem:rbar}
    For all data points $(h_{\IN},h_{\OUT},q_\IN,q_\OUT)$, it holds
    \begin{equation*}
        r_j(x_j,\Delta h,q_\IN,q_\OUT) = 0 \Leftrightarrow 
        \bar r_j(x_j,\Delta h,q_\IN,q_\OUT) = 0.
    \end{equation*}
\end{lem}
\begin{pf}
We have (leaving out function arguments for simplicity),
\begin{align*}
    r_j=0 & \Leftrightarrow 
    U_j[q_{\OUT}-G_{-j}] = \frac{\Delta h -x_jU_j[q_{\IN}-G_{-j}]}{1-x_j} \\
    & \Leftrightarrow \bar r_j = 0,
\end{align*}
where the equivalences follow since $x_j\in(0,1)$ by Proposition~\ref{prop:newprop}, and $U_j$ is uniquely invertible.
    \qed
\end{pf}

\begin{rem} \label{rem:barrj}
    Similar to $\bar r_j$, we can write the residual $r_j$ as $r_j=\Delta h - \widehat{\Delta h} (j,x_j,\Delta h, q_\IN, q_\OUT)$, where $\widehat{\Delta h} (j,x_j,\cdot)$ is the estimated head loss under the assumption that the leak is in pipe~$j$, at position $x_j$. Since the estimate $\widehat{\Delta h}$ depends on all the data $(h_{\IN}, h_{\OUT},q_\IN,q_\OUT)$, and not only  on $(h_{\IN}, h_{\OUT},q_\IN)$, we prefer to proceed with the transformed residual $\bar r_j$ next.
\end{rem}

In order to determine under what perturbations of the data point we cannot refute that pipe~$i$ is leaking at $x_i$, we differentiate $r_i=0$ and obtain  
\begin{multline*}
    dh = x_i U_{\IN,i}'dq_{\IN} - x_i U_{\IN,i}'G_{-i}' dh \\ +(1-x_i)U_{\OUT,i}'dq_{\OUT} -(1-x_i)U_{\OUT,j}'G_{-i}' dh,
\end{multline*}
where $U_{\IN,i}':= U_i'(q_{\IN}-G_{-i}(\Delta h))$, $U_{\OUT,i}':=U_i'(q_{\OUT}-G_{-i}(\Delta h))$, and $G_{-i}':=  G_{-i}'(\Delta h)$. Collecting differentials, we obtain
\begin{multline*}
    [1+G_{-i}'(x_iU_{\IN,i}'+(1-x_i)U_{\OUT,i}']dh \\
   = x_i U_{\IN,i}'dq_{\IN} + (1-x_i) U_{\OUT,i}' dq_{\OUT}.
\end{multline*}
Since $r_i=0$, we have $q_\OUT=\hat q_\OUT(i,x_i,\Delta h,q_\IN)$ and determine the sensitivities of the output flow estimation as
\begin{align*}
    \frac{\partial \hat q_{\OUT}(i,x_i,\cdot)}{\partial q_{\IN}} & =  \frac{dq_{\OUT}(\cdot)|_{dh=0}}{dq_{\IN}} = - \frac{R_{\IN,i}}{R_{\OUT,i}} \\
    \frac{\partial \hat q_{\OUT}(i,x_i,\cdot)}{\partial (\Delta h)} & = \frac{dq_{\OUT}(\cdot)|_{dq_{\IN}=0}}{dh}  = \frac{1+G_{-i}'(R_{\IN,i}+R_{\OUT,i})}{R_{\OUT,i}},
\end{align*}
where we have introduced the pipe section resistances
\begin{align*}
    R_{\IN,i}  := x_iU_{\IN,i}', \quad 
    R_{\OUT,i} := (1-x_i) U_{\OUT,i}'.
\end{align*}

Now we are in a position to study the sensitivity of the residual functions $\bar r_i$, under the assumption that pipe~$k$ is the leaking pipe, meaning $r_k\equiv 0$ and $q_\OUT \equiv \hat q_\OUT(k,x,\Delta h,q_\IN)$. We choose $\Delta 
 h$ and $q_\IN$ as the independent variables (inputs) and $q_\OUT$ as the dependent variable. We then have
\begin{multline*}
    \bar r_i(\Delta h,q_\IN,q_\OUT(\Delta h,q_\IN))  \\ = 
    \hat q_\OUT(k,x,\Delta h,q_\IN) - \hat q_\OUT(i,x_i,\Delta h,q_\IN).
\end{multline*}
and upon differentiation 
\begin{multline}
    d\bar r_i = \frac{\partial \bar r_i}{\partial q_{\IN}}dq_{\IN} +\frac{\partial \bar r_i}{\partial (\Delta h)}dh \\ = \left( \frac{R_{\IN,i}}{R_{\OUT,i}} - \frac{R_{\IN,k}}{R_{\OUT,k}} \right)dq_{\IN} + \left(\frac{1+G_{-k}'(R_{\IN,k}+R_{\OUT,k})}{R_{\OUT,k}} \right.\\ \left.-\frac{1+G_{-i}'(R_{\IN,i}+R_{\OUT,i})}{R_{\OUT,i}} \right)dh. \label{eq:deltar_i}
\end{multline}

Here, we can state a first \emph{negative} result concerning the possibility of pipe isolation, and thus for solving Problem~\ref{problem} for any flows and parallel networks.
\begin{prop}\label{prop:confusion_flow}
    If $\dfrac{R_{\IN,i}}{R_{\OUT,i}} - \dfrac{R_{\IN,k}}{R_{\OUT,k}}\neq 0$, there exists an inflow $q_{\IN}(\Delta h)$ satisfying
\begin{equation}\label{eq:confusion_flow}
    dq_{\IN} =  - \frac{\partial \bar r_i/\partial (\Delta h)}{\partial \bar r_i/\partial q_{\IN}} dh,
\end{equation}
with $\dfrac{\partial \bar r_i}{\partial q_{\IN}}$ and $\dfrac{\partial \bar r_i}{\partial (\Delta h)}$ given in~\eqref{eq:deltar_i},
such that $d\bar r_i = dr_i \equiv  0$,
for all perturbations $dh$.
\end{prop}
\begin{pf}
Assume a flow $q_\IN = q_\IN(\Delta h)$ satisfying~\eqref{eq:confusion_flow}, and insert in~\eqref{eq:deltar_i}.
It follows that $d\bar r_i \equiv 0$ for all $dh$. Finally, we use that $\bar r_i = 0 \Leftrightarrow r_i=0$ from Lemma~\ref{lem:rbar}. \qed
\end{pf}

Hence, if $\partial \bar r_i/\partial q_{\IN}\neq 0$ there always exists a flow~$q_{\IN}(\Delta h)$, which we call a \emph{confusion flow}, such that we cannot reject pipe~$i$ as the leaking pipe. A flow satisfying~\eqref{eq:confusion_flow} may be unlikely in practice, but similar flows yield $\bar r_i \approx 0$ and lead to difficult isolation problems. Also, Proposition~ \ref{prop:confusion_flow} only provides a sufficient condition, and as we shall see in Section~\ref{sec:impossible}, there are \emph{situations when all flows are confusion flows} and Problem~\ref{problem} is inherently ill-posed.
In any case, after the following example, we shall conversely provide network conditions under which \emph{all} flows allow us to reject pipe~$i$, and Problem~\ref{problem} is then surely well-posed.

\begin{exmp}\label{exmp:confusion_flow}
We consider three parallel pipes, where the leak is localized to pipe~1 at $x=0.65$. We assume the head loss functions $U_i(q)=c_i(q|q|+q)$ with $c_1=2$, $c_2=4$, and $c_3=6$. For the computation of the actual flows, we use the leak model $q_\text{leak}=\sqrt{h_\text{leak}}$. In Fig.~\ref{fig:external_flows}, the external flows are shown around the nominal data point $\Delta h = 4.0$ ($h_\IN=5$, $h_\OUT=1$). At this point, we use Proposition~\ref{prop:newprop} to compute the possible leak positions $x=0.65$, $x_2\approx 0.63$, and $x_3\approx 0.64$. Fig.~\ref{fig:confusion_flows_dh4} shows the residual functions around the nominal point. They coincide at $\Delta h=4.0$ since $x_2$ and $x_3$ are computed at this point, but $\bar r_2$ and $\bar r_3$ clearly deviate from zero as we perturb the pressure, and we can reject pipes~2 and 3 as the leaking pipes. In Fig.~\ref{fig:confusion_flows_dh4}, we also see the confusion flows computed around this nominal point. They have been numerically computed as the solutions to $0=\bar r_i(x_i,\Delta h, q_\IN^\text{conf.$i$}(\Delta h),\hat q_\OUT(1,x,\Delta h,q_\IN^\text{conf.$i$}(\Delta h)))$, for $i=1,2$ and varying $\Delta h$. If the flow $q_\IN(\Delta h)$ is replaced with $q_\IN^\text{conf.$i$}(\Delta h)$, then $\bar r_i(\Delta h)$ is forced to zero and we cannot reject pipe~i as the leaking pipe.
\begin{figure}
    \centering
    \includegraphics[width = 1.0\hsize]{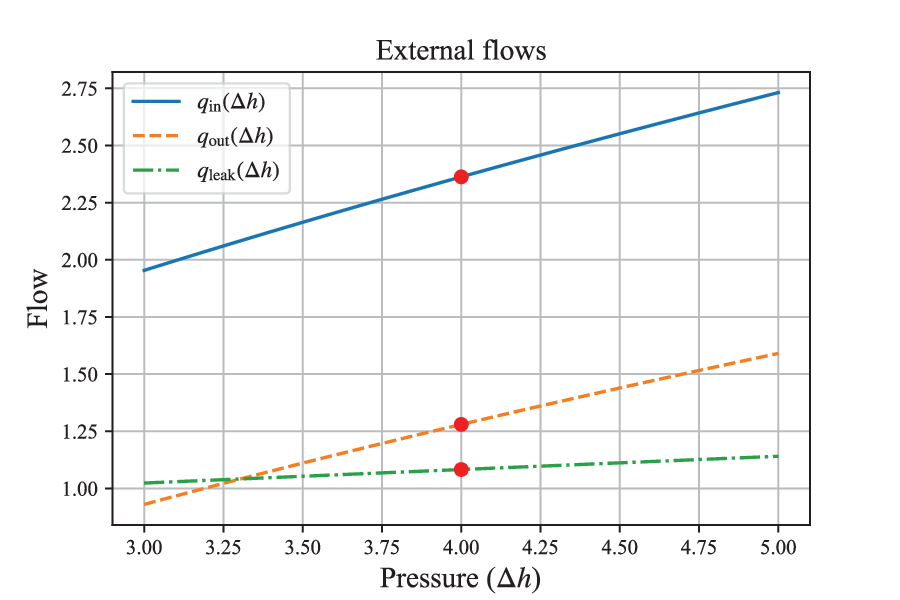}
    \caption{External flows in Example~\ref{exmp:confusion_flow} for nominal value $\Delta h=4.0$.}
    \label{fig:external_flows}
\end{figure} 
\begin{figure}
    \centering
    \includegraphics[width = 1.0\hsize]{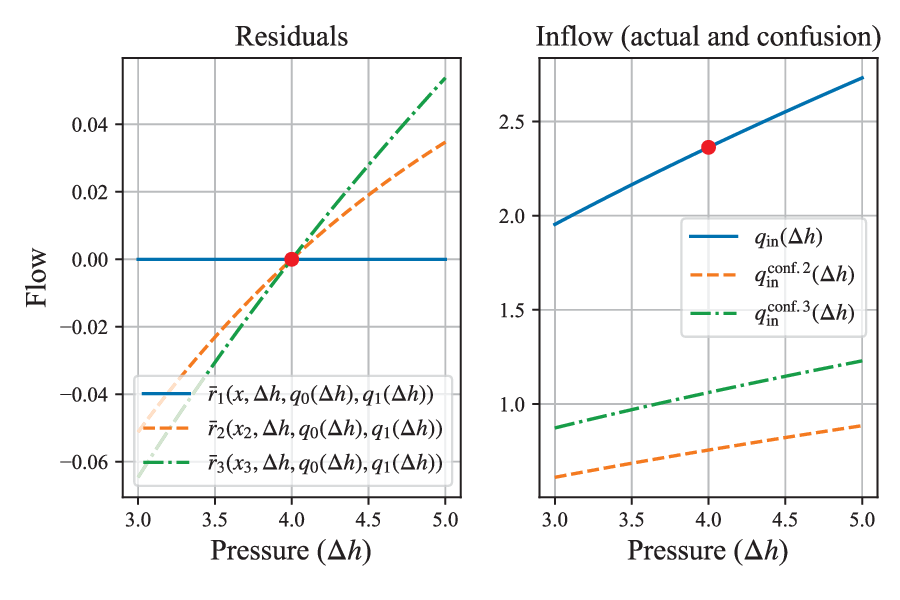}
    \caption{Residuals and confusion flows in Example~\ref{exmp:confusion_flow} for nominal value $\Delta h=4.0$.}
    \label{fig:confusion_flows_dh4}
\end{figure} 

In Fig.~\ref{fig:confusion_flows_dh1}, the residual signals and confusion flows for the nominal data point $\Delta h = 1.0$ ($h_\IN=2$, $h_\OUT=1$) are shown. Proposition~\ref{prop:newprop} now provides leak positions $x=0.65$, $x_2\approx 0.69$, and $x_3\approx 0.72$. There are several noticeable differences to the previous case. First, we note that the magnitudes of the residuals are generally larger, even though the pressure gradient across the pipe system is smaller. Hence, rejecting that pipes~1 and 2 are leaking may be experimentally easier in this case. Second, for $\Delta h \approx 1.0$, all residuals are close to $0$, so a small perturbation may not be enough here. In line with this observation, we can confirm that the confusion flows are almost identical to the actual flow for $\Delta h \approx 1.0$. For larger perturbations, the found confusion flows are noisy due to convergence issues in the numerical equation solver.
\begin{figure}
    \centering
    \includegraphics[width = 1.0\hsize]{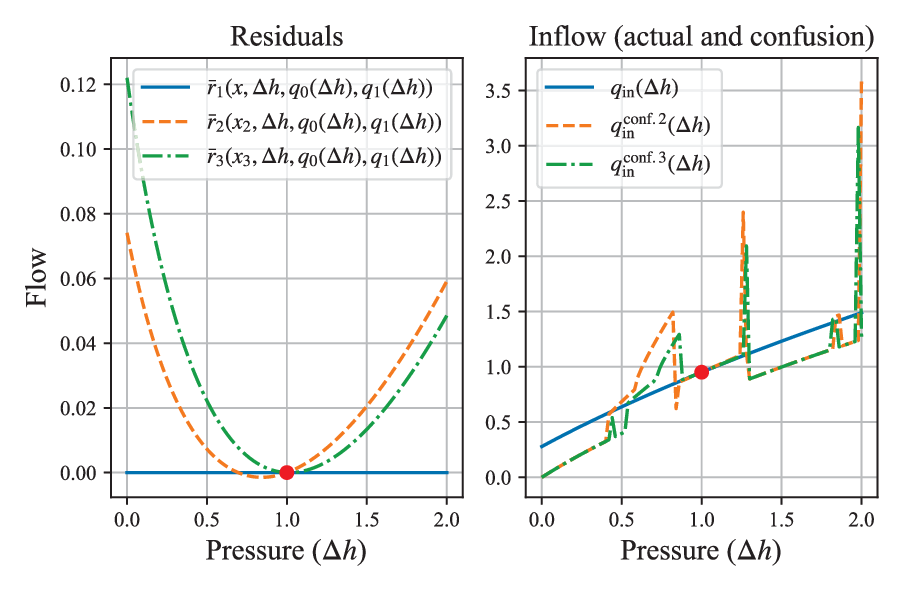}
    \caption{Residuals and confusion flows in Example~\ref{exmp:confusion_flow} for nominal value $\Delta h=1.0$.}
    \label{fig:confusion_flows_dh1}
\end{figure} 
\end{exmp}

Next, we show that there is a generic flow state such that $\partial \bar r_i/\partial q_{\IN} = 0$ and $\partial \bar r_i/\partial (\Delta h)\neq 0$. A \emph{positive} result of~\eqref{eq:deltar_i} is then that any pressure perturbation~$dh\neq 0$ leads to $\bar r_i\neq 0$, and we can correctly reject pipe~$i$.
We propose to consider a nominal state satisfying $\Delta h=0$, and the case when the pipes are of uniform material, as stipulated in the following lemma. Indeed, we saw in Example~\ref{exmp:confusion_flow} that the residual signals were more sensitive for smaller $\Delta h$. We also saw the confusion flow equation was more ill-conditioned, indicating it is hard to find flows that result in small residuals uniformly.

\begin{lem}\label{lem:dHnull} Suppose that $\Delta h=0$ and that the head loss functions satisfy $U_k(q)=c_iU_i(q)$, for all $q$ and some constants $c_i\neq 0$, for all $i\neq k$. Then:
\begin{enumerate}
    \item $x_i=x$; and
    \item $\dfrac{R_{\IN,i}}{R_{\OUT,i}} =\dfrac{R_{\IN,k}}{R_{\OUT,k}}=:\rho$.
\end{enumerate}
\end{lem}
\begin{pf}
\begin{enumerate}
    \item Since $\Delta h=0$, we have for the leaking pipe~$k$ that $0=xU_{k}(q_{\IN})+(1-x)U_{k}(q_{\OUT})$. Note that since $\Delta h=0$, there will only be a non-zero flow through the leaking pipe ($G_{-k}(0)=0$). It follows for any $i\neq k$ that
    \begin{multline*}
        \frac{x}{1-x} = -\frac{U_k(q_{\OUT})}{U_k(q_{\IN})} = -\frac{c_iU_i(q_{\OUT})}{c_iU_i(q_{\IN})}\\ = \frac{x_i}{1-x_i}\Rightarrow x=x_i.
    \end{multline*}
    \item 
    We have for the leaking pipe~$k$ that
    \begin{multline*}
        \frac{R_{\IN,k}}{R_{\OUT,k}}= \frac{x U_k'(q_{\IN})}{(1-x)U_k'(q_{\OUT})} = \frac{x_i U_k'(q_{\IN})}{(1-x_i)U_k'(q_{\OUT})} \\ = \frac{x_i c_i U_i'(q_{\IN})}{(1-x_i)c_i U_i'(q_{\OUT})} = \frac{R_{\IN,i}}{R_{\OUT,i}},
    \end{multline*}
    for any $i\neq k$, since $x=x_i$ (see above) and $U_k'(q)=c_iU_i'(q)$ by the assumption on the loss functions.\qed
    \end{enumerate} 
\end{pf}
\begin{rem}
    The constants $c_i$ can be interpreted as the relative length of pipe~$i$ with respect to pipe~$k$.
\end{rem}
From Lemma~\ref{lem:dHnull}, it directly follows that 
\begin{equation*}
    \frac{\partial \bar r_i}{\partial q_{\IN}} = \frac{R_{\IN,i}}{R_{\OUT,i}} - \frac{R_{\IN,k}}{R_{\OUT,k}} = 0,
\end{equation*}
so that Proposition~\ref{prop:confusion_flow} does not apply. From~\eqref{eq:deltar_i},
\begin{multline*}
    d\bar r_i = \left(\frac{1+G_{-k}'(R_{\IN,k}+R_{\OUT,k})}{R_{\OUT,k}} \right.\\ \left.-\frac{1+G_{-i}'(R_{\IN,i}+R_{\OUT,i})}{R_{\OUT,i}} \right)dh. 
\end{multline*}
We next want to simplify this expression and show when it is surely non-zero. First, we note that in the considered state $\Delta h=0$, there is no flow in the non-leaking pipes, and thus
\begin{equation*}
    G_{-k}'(0) = \sum_{i\neq k} \frac{1}{U_i'(0)},
\end{equation*}
by the inverse function rule.
We define the resistance of the pipe in the non-leaking, zero-flow state as
$R_{0,i} := U_i'(0)$.
\begin{rem}\label{rem:laminar}
For a linear loss function, which is a characteristic of a laminar flow, it holds that
\begin{equation*}
    R_{\IN,i}=x_iR_{0,i}, \quad R_{\OUT,i}=(1-x_i)R_{0,i}, 
\end{equation*}
and therefore $R_{\IN,i}+R_{\OUT,i}=R_{0,i}$. These relations do not hold in general, since the anticipated flows in leaking pipe sections are different and non-zero.
\end{rem}

The following theorem shows that in any nominal state with $\Delta h=0$, any small perturbation in pressure will let us reject that pipe $i\neq k$ is leaking if and only if pipes $i$ and $k$ are not identical (i.e., $U_i\neq U_k$) and the loss functions are not linear (see Remark~\ref{rem:laminar}). Under these network conditions, Problem~\ref{problem} is always well-posed.
\begin{thm}\label{thm:dh0cond}
Suppose that $\Delta h=0$ and that the loss functions satisfy $U_k(q)=c_iU_i(q)$, for all $q$ and some constants $c_i\neq 0$, for all $i\neq k$. 
Then \begin{equation}\label{eq:drifinal}
        d\bar r_i= \left(\frac{1}{R_{\OUT,k}}-\frac{1}{R_{\OUT,i}}\right) \left(1- \frac{R_{\IN,i}+R_{\OUT,i}}{R_{0,i}}\right)dh.
    \end{equation}
In particular, $d\bar r_i \neq 0$ for any $dh\neq 0$, if, and only if, 
\begin{enumerate}
    \item $R_{\OUT,i} \neq R_{\OUT,k}$; and
    \item $R_{\IN,i}+R_{\OUT,i}\neq R_{0,i}$.
\end{enumerate}
\end{thm}
\begin{pf}
    Note that by Lemma~\ref{lem:dHnull} we have
    \begin{equation}\label{eq:rhoident}
        \frac{R_{\IN,k}+R_{\OUT,k}}{R_{\OUT,k}} = \frac{R_{\IN,i}+R_{\OUT,i}}{R_{\OUT,i}} = 1+\rho.
    \end{equation}
    Thus, 
    \begin{equation}\label{eq:preldri}
        \frac{\partial \bar r_i}{\partial (\Delta h)} = \frac{1}{R_{\OUT,k}}-\frac{1}{R_{\OUT,i}} + (1+\rho)(G'_{-k}-G'_{-i}).
    \end{equation}
    Next, we note that
    $G'_{-k}-G'_{-i} = \frac{1}{R_{0,i}}-\frac{1}{R_{0,k}}$,    
    and~\eqref{eq:preldri} can be rewritten as, using~\eqref{eq:rhoident} again,
     \begin{multline}\label{eq:preldri2}
        \frac{\partial \bar r_i}{\partial (\Delta h)} = \frac{1}{R_{\OUT,k}}\left( 1-\frac{R_{\IN,k}+R_{\OUT,k}}{R_{0,k}}\right) \\ + \frac{1}{R_{\OUT,i}}\left( \frac{R_{\IN,i}+R_{\OUT,i}}{R_{0,i}}-1\right).
    \end{multline}
To further simplify the expression, we note that
\begin{align*}
R_{0,k} & = U_k'(0) = c_iU_i'(0) = c_i R_{0,i},   \\
R_{\IN,k}+R_{\OUT,k} & = xU_k'(q_{\IN}) + (1-x)U_k'(q_{\OUT}) \\ 
& = x_ic_iU_i'(q_{\IN}) + (1-x_i)c_iU_i'(q_{\OUT})\\ & = c_i(R_{\IN,i}+R_{\OUT,i}),
\end{align*}
so that $\frac{R_{\IN,k}+R_{\OUT,k}}{R_{0,k}} = \frac{R_{\IN,i}+R_{\OUT,i}}{R_{0,i}}$.
Thus~\eqref{eq:preldri2} simplifies to~\eqref{eq:drifinal},
    which concludes the proof. \qed
\end{pf}

Equation~\eqref{eq:drifinal} is of experimental significance in that it quantifies the sensitivity of the residual~$\bar r_i$ to pipe parameters and flow states. In particular, the first factor shows that when the derivatives $U_{\OUT,k}'(q_\OUT)$ and $U_{\OUT,i}'(q_\OUT)$ are small but different, the sensitivity can be large. The second factor quantifies the nonlinearity of $U_i$ under the assumption that it is the leaking pipe. As the loss function becomes more linear, the sensitivity approaches zero.

\section{Two impossible cases and one possible solution}
\label{sec:impossible}

In Section~\ref{sec:experiment}, we saw that $d\bar{r}_i = 0$ if $R_{out,i} = R_{out,k}$ or $R_{in,i} + R_{out,i} = R_{0,i}$, and hence we were unable to determine whether pipe~$i$ or $k$ were leaking; Problem~\ref{problem} was ill-posed. In this section, we shall see more generally that these cases describe two examples of networks where it is indeed \emph{impossible} to determine, using the model~(\ref{eq:dHnonleak})--(\ref{eq:qout}), which pipe is leaking. In these cases, we can find fixed relative leak positions $x_j$ such that $r_j(x_j, \Delta h, q_{\IN},q_{\OUT}) \equiv 0$ for all possible data points, not only for small perturbations around a nominal point as in Section~\ref{sec:experiment}. 

\subsection{Identical pipes}
The model (\ref{eq:dHnonleak})--(\ref{eq:qout}) summarises all information about a pipe in the head loss function $U_i$. Therefore, if $U_j = U_k$ for some $j$, we can not tell whether pipe $j$ or pipe $k$ is leaking. We state this in Theorem \ref{thm:identical}.
\begin{thm}
    \label{thm:identical}
    If $U_j \equiv U_k$, then $r_j(x_j,\Delta h, q_{\IN},q_{\OUT}) \equiv 0$ for all data points $(h_{\IN},h_{\OUT}, q_{\IN},q_{\OUT})$, and $x_j=x$.
\end{thm}
\begin{pf}
First notice that $G_{-j}(\Delta h) -G_{-k}(\Delta h) = \sum_{i \neq j}U_i^{-1}(\Delta h)-\sum_{i \neq k}U_i^{-1}(\Delta h) = U_k^{-1}(\Delta h)-U_j^{-1}(\Delta h) = 0$. Therefore
$ r_j(x,\Delta h, q_{\IN},q_{\OUT})=\Delta h -xU_j(q_{\IN}-G_{-j}(\Delta h)) -(1-x)U_j(q_{\OUT}-G_{-j}(\Delta h)) = \Delta h - xU_k(q_{\IN}-G_{-k}(\Delta h))-(1-x)U_k(q_{\OUT}-G_{-k}(\Delta h)) = r_k(x,\Delta h, q_{\IN},q_{\OUT}) \equiv 0$.
\qed
\end{pf}
\begin{rem}
    Notice that in Theorem~\ref{thm:identical}, we have $x_j = x$. This means that the measurements make it seem like either pipe $j$ or pipe $k$ could be leaking, in the same relative position $x$.
\end{rem}

\subsection{Linear head loss}
\label{subsec:linear}
Leak isolation is impossible also in the case where the head loss functions are linear for the leaking pipe $k$ and another pipe $j$.

\begin{thm}
    \label{thm:impossiblelinear}
    If pipe $j$ and pipe $k$ both have linear head loss functions $U_j(q) = R_j q$, $U_k(q) = R_k q$, for any constants $R_j,R_k >0$, then for all data points $(h_{\IN},h_{\OUT}, q_\IN, q_\OUT)$ it holds that $r_j(x_j, \Delta h, q_{\IN},q_{\OUT}) \equiv 0$, and $x_j=x$.
\end{thm}
\begin{pf} With linear $U(q) = Rq$, we have $U^{-1}(h) = h/R$. Therefore
\begin{align*}
    r_j(x&, \Delta h, q_{\IN},q_{\OUT}) 
    \\& = \Delta h -xU_j(q_{\IN}-G_{-j}(\Delta h))
    \\
     & \quad -(1-x)U_j(q_{\OUT}-G_{-j}(\Delta h))
    \\ & = \Delta h -xU_j(q_{\IN}-G_{-k}(\Delta h)
    \\ & \quad +U_j^{-1}(\Delta h)-U_k^{-1}(\Delta h))
    \\ &\quad -(1-x)U_j(q_{\OUT}-G_{-k}(\Delta h)
    \\ & \quad +U_j^{-1}(\Delta h)-U_k^{-1}(\Delta h))
    \\ & = \Delta h -xR_j(q_{\IN}-G_{-k}(\Delta h)) -xR_j\left(\dfrac{1}{R_j}-\dfrac{1}{R_k}\right)
    \\ & \quad -(1-x)R_j(q_{\IN}-G_{-k}(\Delta h))
    \\ & \quad-(1-x)R_j\left(\dfrac{1}{R_j}-\dfrac{1}{R_k}\right)
    \\ & = \Delta h\left(1-\dfrac{R_j}{R_j}+\dfrac{R_j}{R_k}\right) 
    \\
    & \quad -\dfrac{R_j}{R_k}xR_k(q_{\IN}-G_{-k}(\Delta h)) 
    \\ & \quad -\dfrac{R_j}{R_k}(1-x)R_k(q_{\OUT}-G_{-k}(\Delta h))
    \\ & = \dfrac{R_j}{R_k} r_k(x, \Delta h, q_{\IN},q_{\OUT}) \equiv 0. 
\end{align*}
\qed
\end{pf}
\begin{rem} Theorem~\ref{thm:impossiblelinear} requires no specification of the other $n-2$ head loss functions, which can be nonlinear. Note also that linear head loss is associated with laminar flows. The result indicates that leak isolation is difficult in pipes carrying laminar flows, even if the pipes have different flow resistance ($R_j \neq R_k$).
\end{rem}

A possible solution in the linear head loss case is to exploit side information about the leak characteristics.

\subsection{Solution to linear loss case via leak function}
\label{sec:leakfunction}
We may add further physical insight to our model to deal with the case of indistinguishable linear head loss pipes. We write
\begin{align}
    \label{eq:Hleakin}
    h_{\IN}-h_{\LEAK} &= xU_k(q_{\IN,k}),
    \\
    \label{eq:Hleakout}
    h_{\LEAK}-h_{\OUT} &= (1-x)U_k(q_{\OUT,k}),
\end{align}
so that~(\ref{eq:dHleak}) is the sum of~(\ref{eq:Hleakin}) and~(\ref{eq:Hleakout}), where~$h_{\LEAK}$ is the hydraulic head at the leak. Equations~(\ref{eq:Hleakin}) and~(\ref{eq:Hleakout}) describe the head loss from the inlet to the leak and from the leak to the outlet, respectively. We augment the model~(\ref{eq:dHnonleak})--(\ref{eq:qout}) with the relation
\begin{align}
    \label{eq:leakfnc}
    q_{\LEAK} = g(h_{\LEAK}).
\end{align}
According to~(\ref{eq:leakfnc}), leakage depends only on the head at the leak position. This is a common assumption, often referred to as \emph{pressure dependent leakage}, which is used, for example, in EPANET \cite{epanet}. 
With linear head loss functions in the leaking pipe~$k$ and the leaking pipe candidate~$j$, as in Subsection~\ref{subsec:linear}, we can solve for the apparent leak hydraulic head in pipe $j$ as a function of the true $h_{\LEAK}$.

\begin{lem}
    \label{lem:leakpressurelinear}
    If $U_j(q) = R_jq$ and $U_k(q) = R_kq$, the apparent leak hydraulic head in pipe $j$ is 
\begin{equation}
    \label{eq:hleakjlemma}
    h_{\LEAK,j} = h_{\LEAK} +(R_k-R_j)x(1-x)q_{\LEAK}.
\end{equation}
\end{lem} 

\begin{pf}
With $j$ being the leaking pipe candidate, we have
\begin{align}
\label{eq:dhinj}
    h_{\IN}-h_{\LEAK,j} &= xR_jq_{\IN,j}\\
    \label{eq:dhoutj}
    h_{\LEAK,j}-h_{\OUT} & = (1-x)R_jq_{\OUT,j}.
\end{align}
Dividing (\ref{eq:dhinj}) by $x$ and (\ref{eq:dhoutj}) by $1-x$ and subtracting, we get 
\begin{align}
\label{eq:hleakj}
    \dfrac{h_{\IN}}{x}-\dfrac{h_{\OUT}}{1-x}-h_{\LEAK,j}\left(\dfrac{1}{x}-\dfrac{1}{1-x}\right) = R_j(q_{\IN,j}-q_{\OUT,j}).
\end{align}
We do the same thing for the truly leaking pipe $k$:
\begin{align}
\label{eq:hleakk}
    \dfrac{h_{\IN}}{x}-\dfrac{h_{\OUT}}{1-x}-h_{\LEAK}\left(\dfrac{1}{x}-\dfrac{1}{1-x}\right) = R_k(q_{\IN,k}-q_{\OUT,k}).
\end{align}
We notice that $q_{\IN,i}-q_{\OUT,i} = q_{\LEAK} = q_{\IN,k}-q_{\OUT,k}$. Subtracting (\ref{eq:hleakj}) from (\ref{eq:hleakk}), we eliminate $h_{\IN}$ and $h_{\OUT}$: $\left(h_{\LEAK,j}-h_{\LEAK}\right)\left(\dfrac{1}{x}-\dfrac{1}{1-x} \right) = (R_{k}-R_j)q_{\LEAK}$. Rearranging gives the result $h_{\LEAK,j} = h_{\LEAK} +x(1-x)(R_k-R_j)q_{\LEAK}.$ \qed
\end{pf}
As a consequence of Lemma~\ref{lem:leakpressurelinear}, if we assume a certain leak function form (the form actually used in EPANET~\cite{epanet}), we can decide which pipe is leaking, if $R_j \neq R_k$. We formalize this result in Theorem \ref{thm:leakfnccontradictionlinear}.
\begin{thm}
    \label{thm:leakfnccontradictionlinear}
    Assuming $U_k(q) = R_kq$, $U_j(q) = R_j q$, $R_j \neq R_k$, and $g(h_{\LEAK}) = C(h_{\LEAK}-h_{yk})^{\beta}$, $0 < \beta \neq 1$, for the elevation level $h_{yj}$, there is no function $g_j(h_{\LEAK,j}) = C_j( h_{\LEAK,j}  -h_{yj})^{\beta_j}$ such that $q_{\LEAK} = g_j(h_{\LEAK,j})$ for all data points $(h_\IN,h_\OUT, q_{\IN}, q_{\OUT})$.
\end{thm}

\begin{pf}
    Substituting $h_{\LEAK}$ in (\ref{eq:hleakjlemma}) by the inverse of $g$, we get $h_{\LEAK,j} =h_{yj} + \left(\dfrac{q_{\LEAK}}{C}\right)^{1/\beta} + x(1-x)(R_k-R_j)q_{\LEAK}.$
    Here $h_{\LEAK,j}$ does not admit the form $h_{\LEAK,j} = h_{yj} +  \left(\dfrac{q_{\LEAK}}{C'}\right)^{1/\beta'}$ for any $C'$,$\beta'$. Hence $q_{\LEAK}$ does not admit the form $q_{\LEAK} = C'(h_{\LEAK,j}- h_{yj})^{\beta'}$, for any $C'$, $\beta'$. \qed
\end{pf}

Theorem~\ref{thm:leakfnccontradictionlinear} says that \emph{if} we can trust the leak to be of the form $q_{\LEAK} = C (h_{\LEAK}-h_y)^{\beta}$, $0< \beta \neq 1$, for some not necessarily known $C$, $\beta$, then there are data points that make it possible to solve the "impossible" linear head loss function case of Subsection~\ref{subsec:linear}. This holds for all $j$ such that $R_j \neq R_k$. However, the difficulty level in rejecting pipe~$j$ as the leaking pipe may vary with the value of $R_j$, as seen in Example~\ref{exmp:leakfcn}.

\begin{exmp}
\label{exmp:leakfcn}
Fig.~\ref{fig:hleakqleak} shows an example with three pipes $1$, $2$ and $3$. Here pipe $U_1(q) = 0.1q$, $U_2(q) = 0.2 q$ and $U_3(q) = 0.3q$. There is a leak in pipe~2 at relative position $x = 0.3$. We also let $h_y = h_{yj} = 0$ so that the hydraulic and pressure heads are equal. We set $C = 50$, $\beta = 0.5$. The plot shows $h_{\LEAK,1}$, $h_{\LEAK,2} = h_{\LEAK}$ and $h_{\LEAK,3}$ as functions of $q_{\LEAK}$. The plot also contains the least squares fit of $H_{\LEAK} = C_jq_{\LEAK,j}^{\beta_j}$, $j = 1,2$. The fit for pipe $1$ contains errors because no $C_1$ and $\beta_1$ fulfill this form; however, the errors are small. In a practical situation, it would still be difficult to tell whether pipe $1$ or $2$ is leaking. On the other hand, $h_{\LEAK,3}< 0 $ for some $q_{\LEAK}$. Therefore we can not fit a function $q_{leak} = C_3 h_{\LEAK,3}^{\beta_3}$ with $C_3, \beta_3 > 0$. The negative $h_{leak,3}$ implies having $3$ as the leaking pipe candidate leads to physically unreasonable behavior, with outflow despite a negative pressure. We conclude that pipe $3$ is not the leaking pipe. 

\begin{figure}
    \centering\includegraphics[scale = 0.6]{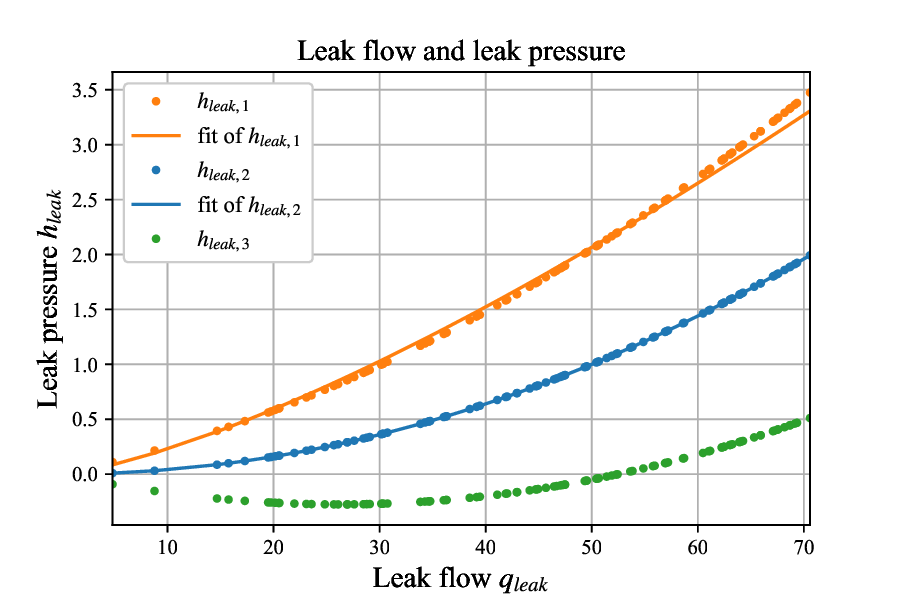}
    \caption{Estimations of $h_{\LEAK,j}$ as function of $q_{\LEAK}$ for three pipes.}
    \label{fig:hleakqleak}
\end{figure}
\end{exmp}

\section{Conclusions}
In this paper, we have approached the water network leak localization problem from a theoretical point of view, formalized in Problem~\ref{problem}. We have analyzed a parallel pipe network structure. Given a set of model assumptions for this structure, we have shown some properties regarding the localizability of leaks. First, we have concluded that the full sensor configuration (two times pressure and flow) is necessary to calculate the leak position. We have provided a formula for the leak position in terms of these sensor measurements. We have shown that one data point is insufficient to tell which of the parallel pipes is leaking. We have determined network conditions under which we can, and cannot, differentiate between leaking pipe candidates given multiple data points. We have also demonstrated that there are certain instances of our model for which it is impossible to isolate the leaking pipe using the given sensor measurements alone. We have shown that, among these, the linear head loss case can sometimes be solved by introducing a leak function. To help display our results, we have provided numerical examples of leak position calculations.

With these efforts, we hope to provide more theoretical understanding of the leak localization problem, which could help in the design of reliable leak localization algorithms. As mentioned, it could also help in the design of challenging leak localization problems for algorithm development and testing. We note also that our results are relevant to larger, more complex networks when they contain parallel pipe subnetworks.

We aim to continue our research by analyzing how uncertainties affect the limitations of leak localization. Given our setting, our results may generalize to other types of potential flow networks, such as electrical circuits and gas pipe networks.

\begin{ack}                               
    We are grateful to the anonymous reviewers for their valuable feedback, which significantly contributed to enhancing the quality of our paper.
\end{ack}

\bibliographystyle{plain}        
\bibliography{autosam}           


\end{document}